# An Ultrasensitive 3D Printed Tactile Sensor for Soft Robotics*

Saeb Mousavi, David Howard, Shuying Wu, Chun Wang

*Abstract*— A new method is presented to manufacture piezo-resistive tactile sensors using fused deposition modelling (FDM) printing technology with two different filaments made of thermoplastic polyurethane (TPU) and polylactic acid-graphene (PLA-G) composite. The sensor shows very high sensitivity (gauge factor~550) and excellent recovery to bending-induced strain and can detect a wide range of pressures. This new technology opens the door for 3D printing soft robotic parts capable of tactile communications.

## I. Introduction

Recently 3D printed tactile sensors have attracted huge attention due to the numerous advantages of additive manufacturing, such as facile integration of different materials into a complex 3D structure [1-3]. One of the materials that is emerging as a suitable candidate for tactile sensing is graphene. Its superior surface area and high conductivity make it ideal for fabrication of conductive polymer composites. Unlike soft physical sensors that were developed using thermoset polymers such as PDMS before, tactile sensors made of thermoplastics are considerably simpler to fabricate and print, because they need no post processing to cure the polymer, and they present a stronger bonding with the nano-structured network embedded inside them due to their considerably higher hardness. Here we present a 3D printed stretchable sensor made of polylactic acid-graphene (PLA-G) conductive polymer composite (CPC) as a piezoresistive sensing material sandwiched between two stretchable thermoplastic polyurethane (TPU) structural layers for acquiring tactile feedback, such as pressure and bending angle. Upon applying stress, which can be originated from an applied force or bending, the graphene network in the PLA matrix will be modified, leading to substantial resistance changes. The design of sandwiched structure with TPU as the structural layers is to increase stretchability. In this work a simple Fused Deposition Modelling (FDM) type printer with two different filaments made of TPU and PLA-G composite is used. The sensor performance is tested for detection of bending angle (bending angles of 0.1 – 26.3 degrees were detected reversibly) and a wide pressure range (from 292 Pa to 487 kPa could be detected reversibly). The current sensor shows very high sensitivity and excellent recovery to bending-induced strain (gauge factor ~ 550) and exhibits very stable response to cyclic pressure and bending. Also, it can distinguish between pressure and bending due to its different sensitivity to these two types of stimuli. The ability to integrate structural and sensing materials into one printed part gives several advantages and bypasses some of the limitations of conventional fabrication methods. This sensor can easily be integrated or attached to soft robotic actuators for acquiring tactile information. Although the 3D printed sensor described here has a simple structure, it demonstrates the potential to create more complex structures and shapes.

## II. Designing The Tactile Sensor

Among conductive polymer composites, PLA-Graphene composite has been selected as the sensing material. Since PLA is not a flexible polymer, we designed only a single layer of PLA-G (thickness ~ 0.2 mm) sandwiched between two layers of TPU (thickness of bottom layer ~ 0.6 mm and thickness of top layer ~ 0.2 mm) to ensure the sensor remains stretchable and bendable. The overall thickness of the sensor is 1 mm.

## III. 3D Printing The Sensor

A standard FDM printer is used for 3D printing the sensor. Our primary goal was to measure the bending angle of soft actuators, so the printing direction was in line with the bending direction. Commercial conductive PLA-G and TPU Ninjaflex filaments (from NinjaTek) were purchased and used to print the sandwich structure. PLA-G layer showed to have a good bonding with TPU in our experiments and no debonding was observed. Figure 1 shows the 3D printed sensor.

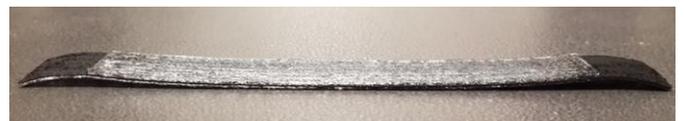

**Figure 1.** The 3D printed sensor. PLA-G is sandwiched between two layers of TPU. At the two ends, PLA-G is designed to be exposed to facilitate wire bonding.

## IV. Tactile Sensing Tests Results

### A. Detection of Bending Angle

The sensor was glued at two ends on an aluminum hinge to test its sensitivity to confined bending. During each experiment, the hinge was bended to a certain degree and returned to its original state rapidly. By measuring the initial gauge length and the radius of curvature, the corresponding strain ($\varepsilon$) induced in the sensor for each bending angle was calculated ($\varepsilon = \Delta L/L_0$), and the gauge factor (GF) was

*Research supported by UNSW and CSIRO.

Saeb Mousavi, Shuying Wu and Chun Wang are with the School of Mechanical and Manufacturing Engineering, UNSW Sydney, NSW, 2033 Australia (Corresponding author's phone: +61421321652; e-mail: s.mousavianchehpoli@unsw.edu.au).

David Howard is with Robotics & Autonomous Systems Group, CSIRO Pullenvale, QLD, Australia (e-mail: david.howard@csiro.au).

calculated subsequently (GF = (ΔR/R$_0$)/ε). The results of the experiments with three different bending angles are shown in Figure 2.

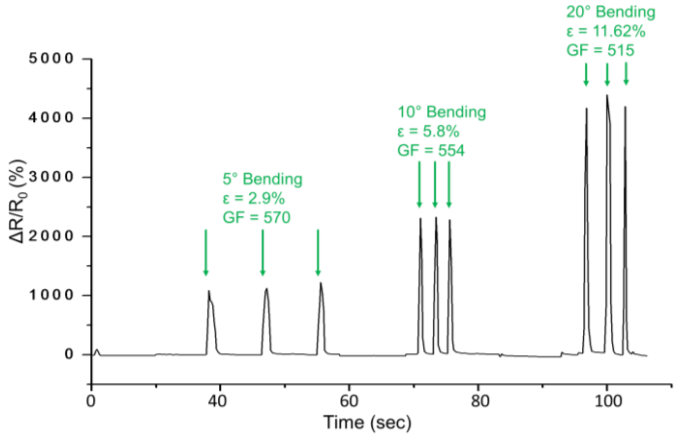

**Figure 2.** Bending angle detection results. The sensor was fixed at two ends on a hinge. The gauge factor (GF) was calculated by calculating the induced strain in each bending cycle.

### B. Contact Pressure Detection

To test the sensor's performance in detecting pressure, Instron 1kN load cell was used to apply periodic contact pressure on the sensor. A small aluminum plate was placed on top of the sensor to uniformly distribute the applied load on the sensor. It was not feasible to measure the smallest detectable pressure using Instron load cell because it was beyond its lowest detectable limit, so this value was calculated. First, the Young's modulus in thickness direction was measured by a strain gauge and then by measuring the strain in thickness direction using Instron, the smallest detectable applied pressure (292 Pa) was calculated. The results of the experiments with three different applied pressures are shown in Figure 3.

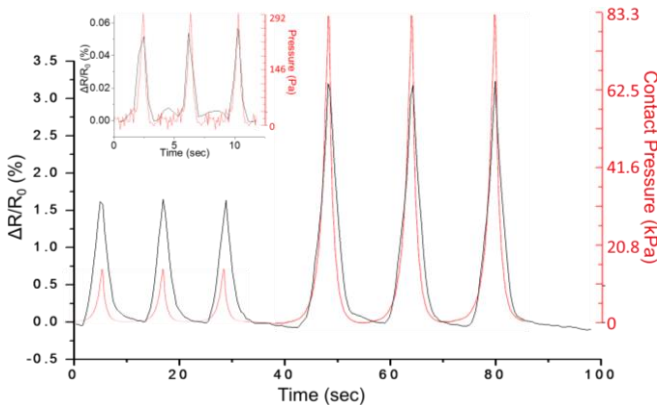

**Figure 3.** Contact pressure detection results. The inset shows the result for the smallest detectable pressure (292 Pa).

## V. CONCLUSION

The present sensor shows an extremely high gauge factor/sensitivity to bending induced strain (GF ~ 550). Its different sensitivity values to bending and pressure helps it to be able to distinguish between pressure and bending stimuli. This sensor was fabricated by a simple 3D printing process. The thermoplastic filaments facilitate the process, because no curing or post-processing is required. Furthermore, this sensor can be printed within or on the surface of soft actuators or can be attached on any surface to give accurate and reliable tactile feedback. The ability to sense contact pressure and bending angle is crucial for a soft actuator and this sensor proved to be a very good candidate to develop such robotic actuators in future.